\documentstyle[12pt,twoside]{article}
{\catcode `\@=11 \global\let\AddToReset=\@addtoreset}
\AddToReset{equation}{section}

\def\greaterthansquiggle{\raise.3ex\hbox{$>$\kern-.75em\lower1ex\hbox{$\sim$}}}
\def\lessthansquiggle{\raise.3ex\hbox{$<$\kern-.75em\lower1ex\hbox{$\sim$}}}

\newcommand{\beq}{\begin{equation}}
\newcommand{\eeq}{\end{equation}}
\newcommand{\beqa}{\begin{eqnarray}}
\newcommand{\eeqa}{\end{eqnarray}}
\newcommand{\beqan}{\begin{eqnarray*}}
\newcommand{\eeqan}{\end{eqnarray*}}
\newcommand{\ba}{\begin{array}}
\newcommand{\ea}{\end{array}}
\newcommand{\no}{\nonumber}

\newcommand{\sgn}{{\rm sgn}}
\newcommand{\sh}{{\rm sh}}
\newcommand{\Det}{{\rm Det}\,}

\newcommand{\ol}{\overline}
\newcommand{\ra}{\rightarrow}

\newcommand{\ve}{\varepsilon}
\newcommand{\vp}{\varphi}

\newcommand{\A}{{\cal A}}

\newcommand{\C}{{\cal C}}

\newcommand{\Ha}{{\cal H}}

\newcommand{\cS}{{\cal S}}

\newcommand{\dint}{\displaystyle \int}
\newcommand{\dsum}{\displaystyle \sum}
\newcommand{\dprod}{\displaystyle \prod}

\newcommand{\dlim}{\displaystyle \lim}

\def\nz{\ifmmode {I\hskip -3pt N} \else {\hbox {$I\hskip -3pt N$}}\fi}
\def\zz{\ifmmode {Z\hskip -4.8pt Z} \else
       {\hbox {$Z\hskip -4.8pt Z$}}\fi}
\def\qz{\ifmmode {Q\hskip -5.0pt\vrule height6.0pt depth 0pt
       \hskip 6pt} \else {\hbox
       {$Q\hskip -5.0pt\vrule height6.0pt depth 0pt\hskip 6pt$}}\fi}
\def\rz{\ifmmode {I\hskip -3pt R} \else {\hbox {$I\hskip -3pt R$}}\fi}
\def\cz{\ifmmode {C\hskip -4.8pt\vrule height5.8pt\hskip 6.3pt} \else
       {\hbox {$C\hskip -4.8pt\vrule height5.8pt\hskip 6.3pt$}}\fi}

\def\au{{\setbox0=\hbox{\lower1.36775ex%
\hbox{''}\kern-.05em}\dp0=.36775ex\hskip0pt\box0}}
\def\ao{{}\kern-.10em\hbox{``}}
\def\lint{\int\limits}
\normalsize
\begin{document}
\bibliographystyle{plain}
\begin{titlepage}
\begin{flushright}
UWThPh-2000-14\\
ESI-864-2000\\
math-ph/0004006\\
April 2, 2000
\end{flushright}

\vspace*{2.5cm}
\begin{center}
{\Large\bf Thermal Correlators\\[10pt]
of Anyons in Two Dimensions%$^\star$
}\\[32pt]

N. Ilieva$^{\ast}$, H. Narnhofer and W. Thirring  \\ [16pt]

{\small\it
Institut f\"ur Theoretische Physik \\ Universit\"at Wien\\
%Boltzmanngasse 5, A-1090 Wien \\
\smallskip
and \\
\smallskip
Erwin Schr\"odinger International Institute\\
for Mathematical Physics\\}

\vfill
\vspace{0.8cm}
\abstract
The anyon fields have trivial $\alpha$-commutator for $\alpha$ not
integer. For integer $\alpha$ the commutators become temperature-dependent
operator valued distributions. The $n$-point functions do not factorize as
for quasifree states.

\vspace{35pt}
%MSC-class: 81T05, 81T40, 82B10, 82B23\\[2pt]
\hspace{1.8cm} $n$-point functions

\end{center}

\vfill
{\footnotesize

%$^\star$ Work supported in part by ``Fonds zur F\"orderung der
%wissenschaftlichen Forschung in \"Osterreich" under grant P11287--PHY;

$^\ast$ On leave from Institute for Nuclear Research and Nuclear Energy,
Bulgarian Academy of Sciences, Boul.Tzarigradsko Chaussee 72, 1784 Sofia,
Bulgaria

%$^\sharp$ E--mail address: ilieva@ap.univie.ac.at
}

\end{titlepage}

\section{Introduction}
The problem of the Bose-Fermi duality in two-dimensional space-time has a
long history. Since the pioneering works in the thirties \cite{J, BNN}
and till now, various of its aspects, conceptual, as well as technical,
remain attractive due to the ambitious idea to extend this remarkable
feature also beyond one space dimension. In recent papers \cite{epj,40},
starting with bare fermions which form the $\,C^*$--algebra $\,\A = {\rm
CAR}(\bf R)$, we have constructed a chain of algebraic inclusions to
substanciate our understanding of this phenomenon:
%close the cycle ``fermions $\ra$ bosons $\ra$
%fermions":
\beq
{\rm CAR}({\it bare}) \subset \pi_\beta(\A)'' \supset\A_c \subset \bar
\A_c \subset \bar\pi_\beta(\bar\A_c)'' \supset {\rm CAR}({\it dressed})
\eeq
To get the first extension, we note that the shift $\,\tau_t\,$ is an
automorphism of $\,\A\,$ which has KMS-states $\,\omega_\beta$ and
associated representations $\pi_\beta$. One then identifies in
$\,\pi_\beta(\A)''\,$ bosonic modes --- the currents, which form the
current algebra $\,\A_c\,$ with a $\beta$-independent structure for $0 <
\beta < \infty$. The crucial ingredient needed at this step is the
appropriately chosen state. We choose the KMS-state which is unique for the
shift over the CAR algebra. Another possibility would be to chose the Dirac
vacuum. This is  what has originally been done in \cite{J,BNN}, in
order to achieve stability for a fermionic system, and recovered later by
Mattis and Lieb \cite{ML} in the context of the Luttinger model.

The essential result is the appearance of an anomalous (Schwinger)
term in the quantum current commutator \beq [j(x),j(x')] =
-\,\frac{i}{2\pi}\,\delta'(x\!-\!x') \eeq based on which
Mandelstam proposed $e^{\,i2\pi \int_{-\infty}^x j(y)dy}$ as
a fermion field \cite{Man}. In this note we construct the anyonic fields
\beq
\Psi_\alpha (x) \simeq e^{\,i 2\pi\sqrt{\alpha}\lint_{-\infty}^{x}j(y)dy}
\eeq
through their action on the bosonic current and study them as
operators in a Hilbert space by exhibiting their n-point function in
a $\tau$-KMS state $\omega$, $\tau$ being the shift automorphism.
This corresponds to taking the crossed product \cite{DHR, BDLR} of
$\A_c$ with an outer automorphism \cite{IN} or, equivalently,
augmenting $\A_c$ by an unitary operator $\,U_\alpha =
e^{\,i2\pi\sqrt{\alpha}j_{\vp_\eta}} \,(\,\alpha \in {\bf R}^+,\,$
and $j_{\vp_\eta}$ being the smeared current with an appropriately
chosen test function $\,\vp_\eta\,$, which converges to a constant
for $\eta\ra 0\,$, $x\ra \infty\,$) to $\,\bar\A_c\,$. Then one discovers
in $\,\bar\pi_\beta(\bar\A_c)''$ anyonic modes which satisfy
Heisenberg's ´Urgleichung´ \cite{Hei}  in a distributional sense
$$
\frac{1}{i} \frac{\partial}{\partial x} \Psi_\alpha(x) =
\pi\sqrt{\alpha} \,[j(x),\Psi_\alpha(x)]_\alpha.
$$
Thus we do not introduce the anyons as fields with deformed
commutation relations as e.g. in \cite{LMP} but we obtain these
relations.

The important point here is that the Hilbert space $\bar\Ha_\beta$
assumes a sectorial structure, being for fixed $\alpha$ a countable
orthogonal sum of sectors with $n$ particles created by $\,U_\alpha$
\beq
\bar\Ha_\beta = \oplus \bar\Ha^n_\beta, \qquad
\bar\Ha^n_\beta = \A_c\,\dprod_{i=1}^{n}
\Psi_\eta(x_i)\vert\Omega\rangle
\eeq

Here the following is to be observed:
\begin{enumerate}
\item $\Psi_\alpha$ as in (1.3) has an infrared and an ultraviolet problem.
The infrared divergence actually shows that admitting the (smeared)
step function as a test function, one creates new elements in the field
algebra which lead to orthogonal sectors in a larger Hilbert space,
Eq.(1.4). The ultraviolet divergence does not lead out of $\bar\pi_\beta$
if we smear $j(y)$ over a region of size $\eta$ to get $\Psi_{\alpha,\eta}$
and consider the renormalized field
$$
\lim_{\eta\ra 0^+}c_\alpha(\eta)\int dx f(x)\Psi_{\alpha,\eta}(x) =
\Psi_\alpha(f) \,
$$
with a suitable $c_\alpha(\eta)$.
This limit exists in a strong sense and $\Psi(f)$ has finite n-point functions.

\item
When the statistical parameter $\alpha$ is an integer, two special families
of such renormalized operators are distinguished: for odd $\alpha$'s
we get fermions and for even $\alpha$'s  --- bosons.
However, only the field $\Psi_1$  turns out  to be a canonical Fermi field,
$$
[\Psi_1^*(x), \Psi_1(x')]_+ = \delta (x-x')\,,
$$
with an n-point function of the familiar determinant form.
$\Psi_2$  is a non-canonical Bose field,
whose commutator is not a $c$-number
$$
[\Psi_2^*(x), \Psi_2(x')] = \delta' (x-x') + ij(x)\delta(x-x')\, .
$$
Similarly, the operator $\Psi_3$
describes a non-canonical  (unbounded) Fermi field. For $\alpha\not\in {\bf Z}$
the anyonic commutator vanishes.
\end{enumerate}

The algebraic chain (1.1) means that the dressed fermions obtained for
special values of $\alpha$ can be
constructed either from bare fermions or directly from the current
algebra, so in this case it cannot be decided whether fermions or bosons
are more fundamental. Moreover, we shall argue that the fermions at
both ``ends" are actually equivalent, so the corresponding algebras
do coincide. To make this statement precise, the correlation functions
arrising in both cases have to be compared and this will be done
subsequently.

\section{The bosonic algebra and its states}

Consider the smeared currents $j(f) = \int j(x)f(x)dx$ with real-valued
functions $f(x) \in \C_0^\infty$, so that neither infrared nor ultraviolet
problems occur. $f$'s form a real pre-Hilbert space to be defined subsequently.
For the bosonic algebra built by the Weyl operators $e^{ij(f)}$,  (1.2)
is replaced by the multiplication law for the unitaries
\beq
e^{ij(f)} \; e^{ij(g)} =
e^{\frac{i}{2} \sigma(g,f)} \; e^{ij(f+g)}\,,
%=e^{i\sigma(g,f)} \; e^{ij_g} \; e^{ij_f}.
\eeq
with a symplectic form %$\sigma(f,g)$
\beq
  \sigma(f,g) =
%\int_{-\infty}^\infty \frac{dp}{(2\pi)^2} \; p \wt f(p) \wt g(-p) =
\frac{1}{4\pi} \int_{-\infty}^\infty dx(f'(x)g(x) - f(x)g'(x)).
\eeq
As an integral kernel the symplectic form reads $\sigma(x-y) = \delta'(x-y)/2\pi$.
Therefore the bosonic algebra satisfies the requirement of a local field theory.

We shall be interested in states that are invariant under a time evolution
$\tau_t$, in particular --- KMS states.
An equilibrium state of a quantum system at a finite temperature  $T =
\beta^{-1}\,$ is characterized by the KMS-condition
$$
\omega_\beta\left(\tau_t(A)\,B\right) = \omega_\beta(B\,\tau_{t+i\beta}A)
$$
with the time evolution $\tau_t$ as an automorphism of the algebra of
observables analytically continued for imaginary times. In the bosonic (current)
algebra $\A_c$ this role is played by the shift
$$
\tau_t j(f(x)) = j(f(x-t))\, .
$$
The time-invariant states are in general defined by the two-point function
$$
\omega(j(f)j(g)) = \int dxdy \,w(x-y)f(x)g(y)\, ,
$$
so for the $\tau$-KMS state at a temperature $\beta=\pi$ the kernel $w(x-y)$ reads
\beq
w(x-y) =-\lim_{\ve\ra0^+}\,\frac{1}{(2\pi)^2\,\sh^2(x-y-i\ve)}\, .
\eeq

If we want to avoid distributions being involved into the calculations, we can
consider the smeared functions
\beq
f_\eta(x) = \int f(y) h_\eta(x-y)dy\, , \quad j(f_\eta) = j_\eta(f)\, ,
\eeq
with some $h_\eta \in \C_0^\infty $. This leads to a new symplectic form
\beq
\sigma(f_\eta, g_\eta) \equiv \sigma_\eta(f,g)\,,
\eeq
$$
\begin{array}{rcl}
\sigma_\eta(x-y) & = & \frac{1}{2\pi}\dint \delta'(z-z')h_\eta(z-x)h_\eta(z'-y)dzdz'\\[6pt]
& = & \frac{1}{2\pi}\dint h_\eta(z-x)h_\eta'(z-y)dz\, .\\[2pt]
\end{array}
$$
In this case $\ve$ of (2.3) need not be finite but just indicates which distribution
is understood. We shall henceforth omit keep writing $\lim_{\ve\ra 0}$.

$\C_0^\infty$ functions and functions with steps are mapped by the smearing
into $\C_0^\infty$ functions. Therefore the bosonic algebra remains
unchanged. The essential point is that the fields $j_\eta(x)$ are not
local any more in the sense that they do not commute if they are
spatially separated. On the other hand, we gain that they become
(unbounded) operators and not only operator-valued distributions.
Their time evolution is given by
$$
\tau_t j_\eta(x)  =  j_\eta(x+t) \, .
$$
Again we can express the state by
$$
\omega (j(f_\eta) j(g_\eta)) = \omega (j_\eta(f)j_\eta(g)) :=
\langle f  \vert g\rangle_\eta
$$
where the corresponding integral kernel
$$
w_\eta(x-y) =  \int w(z-z')h_\eta(x-z)h_\eta(y-z')dzdz'
$$
satisfies
$$
w_\eta(x-y) - w_\eta(y-x) = \sigma_\eta(x-y)\, .
$$

The expectation of the Weyl operators is given by
\beq
\omega(e^{ij_\eta(f)}) = e^{-{1\over 2}\langle f | f \rangle_\eta}\, .
\eeq
Eqs.(2.1),(2.6) imply
\beq
\omega(\prod_k e^{ij_\eta(f_k)}) = \exp{\left\{-{1\over 2}
\left[\dsum_k \langle f_k | f_k \rangle_\eta +
2\dsum_{k<m} \langle f_k | f_m \rangle_\eta \right]\right\}}\, .
\eeq
If  $\lim_{\eta\ra 0} h_\eta(x) = \delta(x)$, then $\lim_{\eta\ra 0}
\langle f | g \rangle_\eta = \langle f | g\rangle $ and the operator $j_\eta(x)$
approaches the distribution $j(x)$.

With the scalar product $\langle f | g\rangle$ the one-particle Hilbert
space $h$ is defined as the closure of the pre-Hilbert space $\C_0^\infty$,
that determines the bosonic von Neumann algebra $\bar\pi_\beta
(\A)''=:\A_B$. Of course,  $\langle f | g\rangle_\eta$ defines the
same Hilbert space, but infrared and ultraviolet divergencies are kept apart.

\section{The extension to anyons}

The field operators we wish to construct in the spirit of Mandelstam are
of the form $\Psi_\alpha (x) \simeq e^{\,i2\pi\sqrt{\alpha}
\int_{-\infty}^{x}j(y)dy} $, so they are defined by the function
$F_x(y) = 2\pi\Theta(x-y)$, whereas the functions from $h$
have to vanish for $x\ra\pm\infty$. The symplectic form (2.2) is
defined for functions that tend to a constant, however they cannot be reached
as limits of functions from $h$. For instance,
\beq
F_{x,\delta}(y) = \Theta(x-y) -
\Theta(x-y-\delta)\, , \quad F_{x,\delta,\eta} \in \C_0^\infty\, .
\eeq
does not work, since $\sigma(F_{x,\delta}, F_{x',\delta'})$ depends on the
order in which the limits $\delta, \delta' \ra\infty$ are taken and only for $\delta=
\delta'\ra\infty$ we get the desired result $i\,\sgn(x-x')$. Since this appears
in the $c$-number part, in no representation can $j(F_{x,\delta})$ converge
strongly. Nevertheless, for functions with the same (nontrivial) asymptotics
at, say, $x\ra\infty$ and whose difference $\in h$ one can succeed  in getting
the expectation values as limits.

The desired extension of the algebra $\A_B$ can be achieved in two
equivalent ways. %Following \cite{CR} or \cite{AMS}
The one-particle Hilbert space can be enlarged \cite{CR, AMS} allowing
also for (appropriately smeared) step functions, e.g. $F_x^\alpha =
\sqrt\alpha F_x$, such that only $F' \in \C_0^\infty$.
For the enlarged algebra the symplectic form (2.2) is kept. If the localization
of the current is given by support $F'$, then fields that are localized in
different regions satisfy the following exchange relation
($\alpha$-commutator)
\beq
e^{ij(F^\alpha)}e^{-ij(G^\alpha)} = e^{i\pi\alpha}
e^{-ij(G^\alpha)}e^{ij(F^\alpha)},
\quad \mbox{supp }F' < \mbox{supp }G'
\eeq
$$
\dlim_{x\ra -\infty}F(x)  =  \dlim_{x\ra -\infty}G(x)  =   2\pi\, ,
$$
$$
\dlim_{x\ra \infty}F(x)  =  \dlim_{x\ra \infty}G(x)  =  0\, .
$$

The same algebra can be obtained by the automorphism $\gamma_{F'}$ of
the initial bosonic algebra $\A_c$
$$
\gamma_{F'} e^{ij(f)} = e^{i\int F'(x)f(x)dx}e^{ij(f)}\, .
$$
This automorphism is not inner but allows the construction of a crossed
product in which $\gamma_{F'}$ is implemented by $e^{ij(F)}$ (compare
\cite{IN}). A given quasifree automorphism $\rho$ on the initial algebra
can be uniquely extended to an automorphism on the enlarged algebra
provided $\gamma_{F'}\rho\gamma_{F'}^{-1}\rho^{-1}$ is an inner
automorphism of  $\A_c$. For instance, $\tau_t$ is extendable
and acts again as a shift on $j(F)$. Also a state on the bosonic algebra
can be extended according to
$$
\omega(e^{ij(F)}) = 0 \qquad \mbox{if } \lim_{x\ra-\infty}F(x) \not= 0\, .
$$
For higher products we combine
$$
\omega(e^{ij(f)} e^{ij(F)} e^{-ij(G)} e^{-ij(g)} ) =
\omega(e^{ij(f)}e^{ij(F-G)} e^{-ij(g)}) e^{i\sigma(F,G)/2} .
$$
The operator $e^{ij(F-G)} \in \A_B$ and the expectation value is well
defined.

As already mentioned, whereas $e^{ij(F)}$ itself cannot be obtained as a
limit of operators in $\A_B$, the situation with the expectation value is
different. We concentrate on the currents $j_\eta(F_x)$, where
%$F_x (y) = \Theta(x-y)$ and
$\eta$ indicates that we have smeared as in (2.4),(2.5) and choose the
test  function (3.1). With this, we get for $\delta > \vert x-\bar x\vert +
\vert \mbox{supp } h_\eta\vert$
\beqan
& & e^{ij_\eta(F^\alpha_{x,\delta})}  e^{-ij_\eta(F^\alpha_{\bar x, \delta})} =
e^{ij_\eta(F^\alpha_{x,\delta}-F^\alpha_{\bar x,\delta}) +
i\alpha\,\sigma_\eta(F_{x,\delta},F_{\bar x,\delta})/2}\\[4pt]
& = & e^{ij_\eta(F^\alpha_x - F^\alpha_{\bar x})}
e^{-ij_\eta(F^\alpha_{x-\delta} - F^\alpha_{\bar x-\delta})}
e^{i\alpha\,\sigma_\eta(F_{x},F_{\bar x})}\, ,
\eeqan
noting that for sufficiently large $\delta$,
$$
\sigma_\eta(F_x - F_{\bar x}, F_{x-\delta} - F_{\bar x -\delta}) =  0
$$
%. Similarly, for $\delta$ large enough
$$
\sigma_\eta(F_x, F_{\bar x}) =
\sigma_\eta(F_{x,\delta}, F_{\bar x,\delta})/2\, .
$$
If we take into account that $\omega$ is translation invariant and
clustering
\beqan
& & \lim_{\delta\ra \infty} \omega (e^{ij_\eta(F^\alpha_x-F^\alpha_{\bar x})}
e^{-ij_\eta(F^\alpha_{x-\delta}-F^\alpha_{\bar x -\delta})}) \\[4pt]
& = & \omega(e^{ij_\eta(F^\alpha_x-F^\alpha_{\bar x})})
\omega(e^{-ij_\eta(F^\alpha_{x-\delta}-F^\alpha_{\bar x-\delta})}) =
e^{-\alpha\langle F_x-F_{\bar x} | F_x-F_{\bar x} \rangle_\eta}\, ,
\eeqan
we can conclude
$$
\omega(e^{ij_\eta(f)}e^{ij_\eta(F^\alpha_x)}e^{-ij_\eta(F^\alpha_{\bar x})}
e^{-ij_\eta(g)}) = \lim_{\delta\ra\infty}
\omega(e^{ij_\eta(f)} e^{ij_\eta(F^\alpha_{x,\delta})/\sqrt{2}}
e^{-ij_\eta(F^\alpha_{\bar x,\delta})/\sqrt{2}} e^{-ij_\eta(g)}) .
$$
Now we can apply (2.7) and obtain
\beqa
& &\omega(\prod_k e^{ij_\eta(F^\alpha_{x_k})}) =
\lim_{\delta\ra\infty}\omega(\prod_k e^{ij_\eta(F^\alpha_{x_k,\delta})/\sqrt{2}})\\
&=& \lim_{\delta\ra\infty}\exp\left\{-{\alpha\over 2}\left
[{1\over 2}\sum_k \langle F_{x_k,\delta} | F_{x_k,\delta}\rangle_\eta
+ \sum_{k<m}\langle F_{x_k,\delta} | F_{x_m,\delta}\rangle_\eta
\right]\right\}\, .\no
\eeqa
We can evaluate the scalar products involved in (3.3):
\beqa
& &\langle F_{x_k,\delta} | F_{x_m,\delta}\rangle_\eta = \int dz
\lint_{-\delta+x_k}^{x_k}dy \lint_{-\delta+x_m}^{x_m} dy'
\frac{\hat h_\eta(z)}{(2\pi)^2\,\sh^2(y-y'-z-i\ve)} \\[6pt]
&=&\frac{1}{(2\pi)^2}\int \hat h_\eta(z) dz \ln\frac{\sh^2(x_k-x_m-z-i\ve)}
{\sh(x_k-x_m+\delta-z-i\ve)\sh(x_k-x_m-\delta-z-i\ve)}\, .\no
\eeqa
To get something finite for $\delta\ra\infty$ we have to take operators of the form
$$
\prod e^{\pm ij(F_{x_k}^\alpha)} := \prod e^{ij(\ol F_{x_k}^\alpha)}\, ,
$$
where $\ol F_{x_k}^\alpha = s_k F_{x_k}^\alpha\,$, $\,s_k = \pm 1$.
The individual expressions diverge with $\delta\ra\infty$. If therefore
the anyon contributions do not neutralize, i.e. if $\sum_k s_k \not=0$,
the expectation value (3.3) vanishes. On the other
hand, if $ \sum_k s_k =0$, we have as many
positive contributions as negative ones, this means for the products
$2r+2r(r-1)$ positive contributions and $2r^2$ negative ones. Those that
contain $\delta$ can be combined in pairs to
$$
 \lim_{\delta\ra\infty}\left[ \ln\frac{\sh(x+\delta-i\ve)}{\sh(y+\delta-i\ve)}
-\ln\frac{\sh(x-\delta-i\ve)}{\sh(y-\delta-i\ve)}\right]
=  \lim_{\delta\ra\infty}\left(\ln\frac{e^{x+\delta}}{e^{y+\delta}} -
\ln\frac{e^{\delta-x}}{e^{\delta-y}}\right) = 0 .
$$
Therefore we remain in the limit $\delta\ra\infty$ with
$$
\omega(\prod_k e^{ij_\eta(\ol F^\alpha_{x_k})}) =
[c^{-2}(\eta)]^{n\alpha}
\exp{\left\{-\int dz \hat h_\eta(z)\alpha\,\ln\left(
\prod_{k<m}\,s_ks_m\,\sh(x_k-x_m-z-i\ve)\right)\right\}}\,,
$$
\beq
c^{-1}(\eta) = \exp{\left[{1\over 2}\int \hat h_\eta(z) \ln\sh(-z-i\ve)dz\right]}\, .
%\quad \lim_{\eta\ra 0}c^\alpha_\ve(\eta) = n_\alpha(\ve)\, ,
\eeq
the latter taking care for the ultraviolet divergence in $\Psi_\alpha$.

In terms of $\Psi$'s this means that the expectation value of a product
of $\Psi$'s and $\Psi^*$'s is different from zero only if there are as many
$\Psi$'s as $\Psi^*$'s, equivalently --- if the ``total" statistic parameter of
creation operators equals the one of annihilation operators, for instance
$\langle \Psi^*_1\Psi^*_1\Psi_4\rangle\not=0$, or otherwise that they lead to
orthogonal sectors of the enlarged Hilbert space.

Performing  the limit $\eta\ra 0$, where $\hat h_\eta(x) \ra \delta(x)$ we achieve our aim ---
construction of the local anyonic field operators. Note that these are now strong
limits of the anyonic Weyl operators as discussed in \cite{CR, AMS, tmp}.

The divergence for $\eta\ra 0$ remains and determines the necessary
renormalization of the operators $\Psi_\alpha$ \cite{tmp}
\beq
\mbox{s-}\lim_{\eta\ra 0}\int f(x)c(\eta)
e^{ij_\eta(F^\alpha_{x})}dx =\int f(x)\Psi_\alpha(x)dx\, .
\eeq

From (3.5), (3.6) we obtain for the $n$-point function
\beqa
&&\omega\left(\Psi_\alpha^*(x_1)\dots\Psi_\alpha^*(x_n)
\Psi_\alpha(y_n)\dots\Psi_\alpha(y_1)\right) \no\\[8pt]
& = & \lim_{\eta\ra 0}[c^2(\eta)]^{n\alpha}
\omega\left( e^{-ij_\eta(  F^\alpha_{x_1} )}\dots e^{-ij_\eta(F^\alpha_{x_n}) }
e^{ij_\eta( F^\alpha_{y_n})}\dots e^{ij_\eta( F^\alpha_{y_1}) }\right) \no \\[6pt]
& = &
\frac{\dprod_{k>l}(\sh(x_k-x_l-i\ve))^\alpha\dprod_{k>l}(\sh(y_k-y_l-i\ve))^\alpha}
{\dprod_{k,l} \left(-2\pi i \,\sh(x_k-y_l-i\ve)\right)^{\alpha}}\,.
\eeqa
The exact exchange relations are hidden in the factor $i^\alpha$ in (3.7) and
we shall return to their detailed analysis later on.

For all $\alpha$'s the two-point function (for $x > x'$ and $\beta =
\pi$)
$$
\langle \Psi_\alpha^*(x) \Psi_\alpha(x') \rangle_\beta =
\langle \Psi_\alpha(x) \Psi_\alpha^*(x') \rangle_\beta =
\left(\frac{i}{2\pi\,\sh(x-x')}\right)^{\alpha} =: S_\alpha(x-x')
$$
has the desired properties
\begin{description}
\item [(i)] {\it Hermiticity:}
$$
S_\alpha^*(x) = S_\alpha(-x) \, \Longleftrightarrow \, \langle
\Psi_\alpha^*(x)\Psi_\alpha(x')\rangle_\beta^* = \langle
\Psi_\alpha^*(x')\Psi_\alpha(x)\rangle_\beta\,;
$$
\item [(ii)] {\it $\alpha$-commutativity:}
$$
S_\alpha(-x) = e^{i\pi\alpha}S_\alpha(x) \, \Longleftrightarrow \,
\langle\Psi_\alpha(x')\Psi_\alpha^*(x)\rangle_\beta =
e^{i\pi\alpha}\langle\Psi_\alpha^*(x)\Psi_\alpha(x')\rangle_\beta \, ;
$$
\item [(iii)] {\it KMS-property:}
\beq
S_\alpha(x) = S_\alpha(-x+i\pi) \, \Longleftrightarrow \,
\langle\Psi_\alpha^*(x)\Psi_\alpha(x')\rangle_\beta =
\langle\Psi_\alpha(x')\Psi_\alpha^*(x+i\pi)\rangle_\beta\, .
\eeq
\end{description}

For $\alpha = 2$ and an arbitrary temperature $\beta^{-1}$
we get like for the $j$'s
$$
\langle\Psi_{2}^*(x)\Psi_{2}(x')\rangle_\beta =
-\frac{1}{\left(2\beta\,\sh{\frac{\pi(x-x'-i\ve)}{\beta}}\right)^2}\, ,
$$
similarly, for $\alpha = 3$ we get a different kind of fermions
$$
\langle\Psi_{3}^*(x)\Psi_{3}(x')\rangle_\beta =
-\frac{i}{\left(2\beta\,\sh{\frac{\pi(x-x'-i\ve)}{\beta}}\right)^3}\, .
$$
These fields, though locally (anti)commuting,  are not canonical and
this becomes transparent by analyzing temperature dependence and
operator structure of their exchange relations.

\section{The $\alpha$-commutator}

As a direct consequence of the Weyl relations (3.2) for the anyonic
Weyl operators (1.3) and with (3.6) in mind, it follows that
$$
\Psi_\alpha^*(x)\Psi_\alpha(y) - e^{i\pi\alpha\,\sgn(x-y)}\Psi_\alpha(y)
\Psi_\alpha^*(x) = 0 \qquad \mbox{for } x\not= y\, .
$$
It remains to calculate the distribution that emerges by bringing
the field arguments together.

Evidently, the case $\alpha \in {\bf Z}$ plays a special role since
we then deal with Fermi- or Bose- commutation relations. We first
concentrate on $\alpha = 1$. Then we observe
$$
\omega(\Psi^*(x_1)\dots\Psi(y_n)) = \Det \omega(\Psi^*(x_i)\Psi(y_k))
= \Det w(x_i-y_k)\, .
$$

\noindent {\bf Proof:} From Cauchy's determinant formula
$$
\Det (x_i-y_k)^{-1} = \frac{\dprod_{i>k}(x_i-x_k)\dprod_{i>k}(y_i-y_k)}
{\dprod_{i,k}(x_i-y_k)}
$$
one gets
$$
\Det \frac{\sqrt{x_iy_j}}{(x_i-y_j)} =
\dprod \sqrt{x_i}\dprod \sqrt{ y_j}\,\,\frac{\dprod_{i>j}(x_i-x_j)\dprod_{i>j}(y_i-y_j)}
{\dprod_{i,j}(x_i-y_j)} = \Det \frac{1}{\sqrt{x_i/y_j}-\sqrt{y_j/x_i}}\, .
$$
This corresponds by replacing $\sqrt x$ by $e^x$, $\sqrt y$ by $e^y$, to
\beq
\Det\frac{1}{\sh(x_i-y_k-i\ve)} = \frac{\dprod_{i>k}
\sh(x_i-x_k-i\ve)\dprod_{i>k}\sh(y_i-y_k-i\ve)}
{\dprod_{i,k}\sh(x_i-y_k-i\ve)}\, .
\eeq

\medskip
\noindent{\bf Theorem}

For $\alpha = 1$ the renormalized fields $\Psi_1^*(x), \Psi_1(x)$ canonically
anticommute
$$
\Psi_1^*(x)\Psi_1(y) + \Psi_1(y)\Psi_1^*(x) = \delta(x-y)\, .
$$

The state over the algebra of fermions is the quasifree state given
by the two-point function
$$
\omega(\Psi_1^*(x)\Psi_1(y)) = \frac{i}{2\pi\,\sh(x-y-i\ve)}\, .
$$
It satisfies the KMS condition with respect to the shift for temperature
$\beta=\pi$ (comp. (3.8)). For arbitrary temperature by scaling arguments
it follows
$$
\omega_\beta(\Psi^*_1(x)\Psi_{1}(y)) =
\frac{i}{2\beta\,\sh \frac{\pi(x-x'-i\ve)}{\beta}}\, ,
$$
with the same commutation relations.

For arbitrary $\alpha$ we have to analyse in detail the expectation
\beqa
\omega_\beta(\Psi^*_\alpha(x_1)\dots\Psi_\alpha(y_n)) & = &
\frac{\dprod_{k>l}[\sh(x_k-x_l-i\ve)]^\alpha
\dprod_{k>l}[\sh(y_k-y_l-i\ve)]^\alpha }
{\dprod_{i,j}\sh(x_i-y_j-i\ve)^\alpha}\no \\[6pt]
& = &\left( \Det\,\sh^{-1}(x_i-y_j-i\ve)\right)^\alpha\, .
\eeqa
Evidently, for $\alpha\not=1$ the state is determined again by the
two-point function but not in a way that corresponds to a truncation.
In order to deduce the commutation relations it is preferable to
evaluate
\beqan
&&\int \omega\left(e^{ij(f_1)}\left[ \Psi^*_\alpha(x)\Psi_\alpha(y) -
e^{i\pi\alpha\,\sgn(x-y)}\Psi_\alpha(y)\Psi^*_\alpha(x)\right] e^{-ij(f_2)}\right)
g_1(x)g_2(y)dxdy \\[8pt]
&=&\int \omega(e^{ij(f_1)} e^{-ij(f_2)}) \left(\frac{1}{\sh^\alpha (x-y-i\ve)} -
\frac{(-1)^\alpha}{\sh^\alpha(y-x-i\ve)}\right)(i/2\pi)^\alpha\\[8pt]
&\times&e^{2\pi\sqrt{\alpha} [\langle f_1 | \Theta(x)\rangle -
\langle f_1 |\Theta(y)\rangle -\langle \Theta(x) | f_2\rangle
+\langle \Theta(y) | f_2\rangle]}\, g_1(x)g_2(y) dx dy\,.
\eeqan
Therefore the distribution to be considered is
$$
\lim_{\ve\ra 0^+}\left(\frac{1}{\sh^\alpha (x-i\ve)} -
\frac{1}{\sh^\alpha(x+i\ve)}\right)\, , \quad \alpha\not\in {\bf Z}^+\, .
$$
This expression tends to 0 for $x\not=0$. As for the singularity at $x=0$,
after a partial integration
one realizes  that it suffices to assume $0<\alpha<1$ and to
evaluate the remaining integral around the origin, so with
$´\sh(x-i\ve)^{-\alpha}\sim (x-i\ve)^{-\alpha}$
\beqan
& &\lim_{\ve\ra 0} \lint_{-\delta}^\delta
\left(\frac{1}{(x-i\ve)^\alpha} -
\frac{1}{(x+i\ve)^\alpha}\right) f(x) dx \\[6pt]
& = &\lim_{\ve\ra 0}\lint_{-\delta/\ve}^{\delta/\ve}
\left(\frac{1}{(y-i)^\alpha} -
\frac{1}{(y+i)^\alpha}\right) f(\ve y) \ve^{1-\alpha}dy  = 0
\eeqan
Therefore for anyonic fields with noninteger statistic parameter the
$\alpha$-commutator vanishes
$$
\Psi_\alpha^*(x)\Psi_\alpha(y) - e^{i\pi\alpha\,\sgn(x-y)}\Psi_\alpha(y)
\Psi_\alpha^*(x) = 0 \qquad \alpha\not\in{\bf Z}^+\, .
$$

For $\alpha$ integer we can perform again partial integration. This
gives us for $\alpha=2$
$$
\lim_{\ve\ra 0} \left(\frac{1}{\sh^2(x-i\ve)} -
\frac{1}{\sh^2(x+i\ve)}\right)= -2\pi i \delta'(x)\, .
$$
Therefore
\beqan
&&\int\,\omega\left(e^{ij(f_1)}[\Psi_2^*(x)\Psi_2(y) -
\Psi_2(y)\Psi_2^*(x)]e^{-ij(f_2)}g_1(x)g_2(y)\right)\,dx\,dy\\[6pt]
&=&{i\over 2\pi}\int \delta'(x-y)\omega(e^{ij(f_1)}e^{-ij(f_2)})
e^{2\pi\sqrt{2}[\langle f_1 | \Theta(x)\rangle -
\langle f_1 |\Theta(y)\rangle -\langle \Theta(x) | f_2\rangle
+\langle \Theta(y) | f_2\rangle]}
\\[4pt]
&&\times g_1(x)g_2(y)\,dx\,dy +  i\sqrt 2 \int g_1(x)g_2(x)\omega(e^{ij(f_1)}e^{-ij(f_2)})\,dx
\eeqan
or
$$
[\Psi_2^*(x), \Psi_2(y)] ={i\over 2\pi} \delta'(x-y) +i\sqrt 2\delta(x-y)j(x)\, .
$$
This follows from comparing the contributions in the expectation
value
$$
\delta'(x-y)\,e^{[\langle f_1 | \Theta(x)\rangle -
\langle f_1 |\Theta(y)\rangle -\langle \Theta(x) | f_2\rangle
+\langle \Theta(y) | f_2\rangle]}\,%\\[4pt]
\longrightarrow \,\delta(x-y)[\langle f_1 | \delta(x)\rangle - \langle \delta(x) | f_2\rangle]
$$
with
$$
\omega(e^{ij(f_1)}j(x)e^{-ij(f_2)}) =
{d\over d\gamma}\,\omega(e^{ij(f_1)}e^{-ij(f_2)})\,
e^{\gamma[\langle f_1 |\delta(x)\rangle - \langle \delta(x)| f_2\rangle]}.
$$

Thus the bosonic fields of the first level do not satisfy canonical
commutation relations. Though the two-point functions look similar,
for the initial bosonic algebra this was a two-point function of currents
whereas now it is a two-point function of fields, that are not invariant
under gauge automorphisms but are adjoint of each other.

For $\alpha=3$ again fermionic fields are obtained. Partial integration
yields (we do not fix the temperature)
\beqan
&&\omega\left(e^{ij(f_1)}[\Psi_3^*(x)\Psi_3(y) +
\Psi_3(y)\Psi_3^*(x)]e^{-ij(f_2)}\right) \\[6pt] %g_1(x)g_2(y)\right)\\[6pt]
&=&-\frac{1}{8\pi^2}\left(\delta''(x-y) -
\frac{\pi^2}{\beta^2}\delta(x-y)\right)\,
e^{2\pi\sqrt 3 [\langle f_1 | \Theta(x)\rangle -
\langle f_1 |\Theta(y)\rangle -\langle \Theta(x) | f_2\rangle
+\langle \Theta(y) | f_2\rangle]}\, .
\eeqan
Here the $\delta''(x-y)$ splits again in a distribution acting only on the
smearing function of the fermion fields and additional terms, so that
we can write the commutator as an operator-valued distribution
$$
[\Psi_3^*(x), \Psi_3(y)]_+ = -{1\over 8\pi^2}[\delta''(x-y) -2\pi\sqrt 3
\delta'(x-y)j(x) + 12 \pi^2\delta(x-y)j^2(x)] + {1\over 8\beta^2}\delta(x-y)
$$
or more precisely,
\beqan
[\Psi_3^*(f), \Psi_3(g)]_+ &=& -{1\over 8\pi^2}\int dx \left\{ {d^2\over dx^2}
 f(x)g(x) -  2\pi\sqrt 3 j(x)\left({d\over dx} f(x)g(x)\right)\right.\\[6pt]
&+& \left. 12\pi^2 f(x)g(x)j^2(x) - (\pi/\beta)^2 f(x)g(x)\right\}\, .
\eeqan

Thus we are faced with a new feature: the commutation relations become
temperature dependent, though they remain local. The operators $\Psi_3(f)$ are
obviously not bounded. The fermions of the third level form
a subalgebra of the whole fermionic algebra (the fermions of the first level).
This demonstrates that representations corresponding to different
temperatures are inequivalent not only globally, but already also locally.

\vspace{0.5cm}

\noindent
{\bf Back to the current algebra}

\smallskip
\noindent
The current algebra can be considered as the subalgebra of any anyonic
algebra $\{\Psi_\alpha(x)\}''$ that is invariant under the gauge automorphism
$\gamma_s\Psi_\alpha(x) = e^{i\alpha s}\Psi_\alpha(x)$. Furthermore, the local
fields $j(x)$ can be directly constructed  out of the anyonic fields $\Psi_\alpha(x)$.
Starting with the fermions $\Psi_1(x)$ this procedure is well known \cite{J, BNN}. In
fact, $\Psi^*(x)\Psi(x)$ is at best a quadratic form and by point splitting it becomes
the operator-valued distribution \linebreak $\Psi^*(x+\ve)\Psi(x-\ve)$. We can relate this to
the bounded operator $e^{ij(f_{x,\ve})}$ with $f_{x,\ve}(y)=1$ for $x-\ve<y<x+\ve$,
otherwise $0$. But then we produce by the sharp edge  of  $f$ an ultraviolet problem
with diverging expectation values. To start with a well defined expression with a zero
expectation value we consider
\beq
j(x) = \lim_{f_{x,\ve}(x)=1\atop\int f_{x,\ve}(y)dy {\longrightarrow\atop \ve\ra 0} 0}
\frac{e^{i\alpha j(f_{x,\ve})} - e^{-i\alpha j(f_{x,\ve})}}
{2i\alpha \int f_{x,\ve}(y) dy}\, .
\eeq
Now we choose
$$
f_{x,\ve}(y) := \left\{\ba{cl} 0 & \,\,\mbox{ for }  y<x -\ve, \,\, y>x+\ve\, , \\[4pt]
1-\left \vert {x-y\over \ve}\right\vert &\,\, \mbox{ for } x- \ve <y< x+\ve\, .  \ea \right.
$$
This can be interpreted as a smearing in the sense of (2.4) so that
$$
e^{i\alpha j(f_{x,\ve})} = \Psi^*_{\alpha,\ve}(x+\ve)\Psi_{\alpha,\ve}(x-\ve)\, ,
$$
whereas
$$
e^{-i\alpha j(f_{x,\ve})} = \Psi^*_{\alpha,\ve}(x-\ve)\Psi_{\alpha,\ve}(x+\ve)
$$
and $\int f_{x,\ve}(y) dy$ corresponds to the renormalization factor needed
to pass from the anyonic Weyl operators to the anyonic fields. Of course, $j(x)$ is
not an operator but has to be smeared by a suitable $f$. Then (4.3) becomes a
strong resolvent limit. In a sense we thus provide a replacement of the familiar
expression
$$
j(x) = \,:\Psi^*(x)\Psi(x):
$$
that has been used to construct  a bosonic current algebra with
Schwinger commutation relations out of the fermions. In fact, since we obtain
the anyonic fields as limit of anyonic Weyl  operators that are not quadratic in
the Fermi field, the current emerges out of a kind of a Dirac sea filling.
This procedure is state dependent and the states under consideration are not
locally normal, therefore the structure of the Dirac sea,
the higher-level fields exhibit a nontrivial temperature
dependence. Nevertheless, the structure of the currents defined by (4.3)
is $\alpha$ and $\beta$ independent.

\section{The Luttinger model}
The Luttinger model has been designed to describe a one-dimensional
interacting electron system. The model spectrum in the ground state
consists of plasmons with a well-defined energy \cite{ML}. The essential
input in these considerations is the fact that the free part of the Hamiltonian
makes a Bogoliubov transformation necessary, so that the Bose operators,
which replace (appropriately) the products of fermionic creation and
annihilation operators $a^*(x)a(x)$, satisfy the Schwinger commutation
relations (1.2).

Several interaction potentials can be considered

\vspace{6pt}
\begin{enumerate}
\item $H=H_0 + \lambda\int j_1(x)v(x-y)j_2(y)$;\\
\item $H=H_0 + \lambda\int(j_1(x)+j_2(x))v(x-y)(j_1(y)+y_2(y))$;\\
\item $H=H_0 + \lambda\int j_1(x)v(x-y)j_1(y)$.\\
\end{enumerate}

With $\rho_i(p)=\int j_i(x)e^{ipx}dp$, (1.2) leads to

\vspace{8pt}
\begin{enumerate}
\item $(\ddot\rho_1(p)+\ddot\rho_2(p)) = -p^2(1-\lambda^2\tilde v^2(p))
(\rho_1(p)+\rho_2(p))$;\\
\item $(\ddot\rho_1(p)\pm\ddot\rho_2(p)) = (p^2+2\lambda p \tilde v(p))
(\rho_1(p)\pm\rho_2(p))$;\\
\item $\dot\rho(p) = p(1+\lambda \tilde v(p))\rho(p)$.\\
\end{enumerate}

In all cases we obtain a quasifree evolution for the currents. There exist
the corresponding KMS-states, provided they are positiv definite (which
might bring in some restrictions on the coupling constant).  With
$\tau_tj(f)=j(e^{i\ve(p)t}f)$,
\beqan
\omega(j(f)j(g)) &=& \langle f\vert g\rangle_{\beta,v} =
\int f(x)g(y)w(x-y)_{\beta,v} \\[6pt]
&=& \lint_{-\infty}^\infty
\frac{p}{1-e^{-\beta\ve(p)}}\tilde f(p)\tilde g(-p)dp \\[6pt]
&=& \lint_0^\infty \frac{p}{1-e^{-\beta\ve(p)}}
(\tilde f(p) \tilde g(-p)+\tilde f(-p)\tilde g(p)e^{-\beta\ve(p)})\, .
\eeqan

As for the free evolution, the state can be extended to the anyonic
Weyl operators as in (3.3), resulting in a KMS-state for the extended
algebra. For sufficiently smooth short-range potentials
$$
\lim_{p\ra\infty}{\ve(p)\over p} = 1, \qquad \lim_{x\ra\pm\infty}
{w(x)_{\beta,v}\over w(x)_\beta} = 1\, .
$$
Therefore the infrared behaviour in $\langle F_{x,\delta}\vert
F_{\bar x,\delta}\rangle$ for $\delta\ra\infty$ can be controled as
for the free case. Since $\tilde w_\beta(p)-\tilde w_{\beta,v}(p) \in \cS$
also the ultraviolet behaviour is unchanged. It is essential again that
the bosonic one-particle Hamiltonian is positiv definite and does not
produce an additional singularity at some $p\not=0$. Also here the
$n$-point functions become products of two-point functions (4.2).
However, already for the $\alpha=1$ fermions  $\ve(p)\not=p$ and they
do not allow the decomposition (4.1), so that the state over the fermionic
algebra is not quasifree any more. Nevertheless, the resulting fermions
satisfy canonical anticommutation relations because the leading
singularity in $w_{\beta,v}(x)$ coincides with that of $w_\beta(x)$
and determines the commutation relations
$$
\omega_{\beta,v}\left(e^{ij(f_1)}[\Psi^*(x)\Psi(y)+\Psi(y)\Psi^*(x)]
e^{-ij(f_2)}\right) = \omega_{\beta,v}(e^{ij(f_1)}e^{-ij(f_2)})\delta(x-y)\, .
$$
This shows that $[\,.,.\,]_+$ is a $c$-number and that $\omega_{\beta,v}$
is in fact a KMS-state for fermions with a point pair interaction, provided
we accept that it results from a  regularization procedure involving both
delocalization and filling of the Dirac sea.
Due to the fact that only the leading singularity remains unchanged,
one should be aware of the indispensable dependence of the exchange
relations of the anyonic fields (especially of the higher-level Bose- and
Fermi- fields) both on the temperature and interaction.

\section{Conclusions}
We have represented the two-dimensional chiral anyonic field $\Psi_\alpha$
as an operator in a Hilbert space and have studied its exchange relations
and its thermal correlation functions. The latter are still determined by the
two-point function
$$
\omega_\beta(\Psi^*_\alpha(x)\Psi_\alpha(y)) =
\left(\frac{i}{2\beta\,\sh \frac{\pi(x-x'-i\ve)}{\beta}}\right)^\alpha ,
$$
however not via simple truncation, but as
$$
\omega_\beta(\Psi^*_\alpha(x_1)\dots\Psi_\alpha(y_n)) =
\left( \Det \frac{i}{2\beta\,\sh \frac{\pi(x_i-y_k-i\ve)}{\beta}}\right)^\alpha\, .
$$

For $\alpha$ not integer the $\alpha$-commutator vanishes, for $\alpha$ odd
the fields obtained are fermions and for $\alpha$ even they are bosons.
However, for $\alpha>1$ the fields still being local, are not canonical, as it
follows from the operator structure of their (anti)commutators. Moreover,
already the first noncanonical fermions are not bounded and their commutation
relations exhibit essential temperature dependence, so representations
corresponding to different temperatures are not locally normal. Along this line
of considerations, in the Luttinger model we have found an example of a not
quasi-free KMS-state.

\section{Acknowledgments}

We are grateful to A. Alekseev and I. Todorov for stimulating our interest
into the problem and to H.J. Borchers, E. Lieb and B. Schroer for suggestive
discussions .

N.I. thanks the International Erwin Schr\"odinger Institute for
Mathematical Physics where the research has been performed, for
hospitality and financial support. This work has been supported
in part also by ``Fonds zur F\"orderung der wissenschaftlichen
Forschung in \"Osterreich" under grant P11287--PHY.

\end{document}